\documentclass[prd,preprint,aps,amsfonts,amssymb,nofootinbib,tightenlines]{revtex4}
\usepackage[dvips]{color}                 
\usepackage{graphicx}

\input{epsf.sty}

\def\b\mu{{\bf \mu}}

\begin{document}

\title{{\Large {\bf Infrared Behavior of the Pressure in $g \phi^3$ Theory Reexamined}}}

\author{Heron Caldas \footnote{Email address: hcaldas@ufsj.edu.br} and A.L. Mota \footnote{Email address: motaal@ufsj.edu.br}}
\affiliation{Universidade Federal de S\~{a}o Jo\~{a}o del Rei, UFSJ, Departamento de Ci\^{e}ncias Naturais, DCNAT, Caixa Postal 110, S\~{a}o Jo\~{a}o del Rei, CEP: 36301-160, MG, Brazil}
\date{August, 2004}

\begin{abstract}
We reinvestigate the infrared behavior of the pressure in the $g \phi^3$ scalar theory in six dimensions. This problem was first studied by Almeida and Frenkel and more recently by Carrington et al., that certified their results under certain approximations. We employ an alternative technique, instead of the approximation methods necessary to truncate the Schwinger-Dyson equations, often considered to calculate the pressure nonperturbatively. A daisy-type sum, implemented through the modified self-consistent resummation (MSCR), is enough to take care of the infrared divergences ensuring the finiteness of the pressure. 
\end{abstract}

\maketitle                          

\vspace{0.5cm}

\newpage

\section{Introduction} 

The physical problems which arise in ultrarelativistic heavy ion collisions or in the 
study of the very early Universe require a consistent treatment of field theory at finite temperature (FTFT)~\cite{Kapusta}. In the context of FTFT one of the most fundamental objects is the partition function, from which all information concerning the equilibrium thermodynamic macroscopic properties of the system can be obtained. One of such thermodynamic properties is the pressure. However, it was noted by Linde~\cite{Linde} that the thermodynamic pressure of the Yang-Mills theory cannot be calculated perturbatively beyond fifth order in the coupling constant. This is due to the infrared (IR) divergences which emerge in FTFT. Various nonperturbative approaches has been used in the last decades trying to circumvent the IR problem cited above~\cite{Almeida,Drummond,Carrington}. Among these nonperturbative methods, are the ones based on the Schwinger-Dyson equations, the hard thermal loop resummation scheme and others. 

In this paper we employ an alternative nonperturbative method, the modified self-consistent resummation (MSCR), which we have developed recently~\cite{Caldas1,Caldas2,Caldas3,Caldas4}, to study the infrared behavior of the pressure in the $g \phi^3$ scalar theory in six dimensions. This model has been chosen due to its similarities to the Yang-Mills theory. This problem was firstly studied by Almeida and Frenkel~\cite{Almeida} through the Schwinger-Dyson equations in the ladder approximation and more recently by Carrington et al.~\cite{Carrington}, that certified (qualitatively) their results under certain approximations. The MSCR resums higher-order terms in a nonperturbative way curing the problem of breakdown of the perturbative expansion. Another advantage of the MSCR is that one can sum an infinite subset of diagrams in order to cancel the IR divergences~\cite{Caldas1,Caldas2,Caldas3,Caldas4}. We show in this work that only one recalculation (resummation) of the self-energy, is enough to obtain an IR divergent free result. We also obtain an analytic (finite) expression for the pressure which was computed in two approximations.

The paper is organized as follows. In Sec. II we give the details of our approach, which is based on the MSCR formalism and derive an expression for the pressure. In Sec. III we compare the results obtained by using two approximations to compute the self-energy. We conclude the paper in Sec. IV.

\section{A Nonperturbative approach}

In this section we perform the resummation of an infinite subset of diagrams employing the MSCR. To one loop order, the self-energy for the massless $g \phi^3$ theory in six dimensions is given by

\begin{equation}
\label{eq1}
\Pi(k_0,k)=\frac{1}{2}\frac{g^2 T}{(2 \pi)^5} \sum_{n} \int_{\lambda}^{T} d^5 p~G(p)~G(p + k),
\end{equation}
where $g$ is a positive coupling constant, $\lambda$ is an infrared cutoff, $T$ is an ultraviolet cutoff on the momentum integration and $G(p)=\frac{1}{w_n^2+p^2}$ is the free massless particle propagator, with $w_n=2n \pi T$ and $k_0=i w_l$. As we are interested in study the infrared behavior of the theory, we take only the zero frequency ($n=0$) terms in the equation above. 
To solve Eq.~(\ref{eq1}), we use the maximization procedure adopted by Almeida and Frenkel \cite{Almeida}, 
\begin{equation}
\label{eq2}
({\bf p}+{\bf k})^2 \equiv \cases{p^2 ~{\rm for}~ p>k, \cr k^2 ~{\rm for}~ k>p.}
\end{equation}
In the limit $\lambda \to 0$ and at leading order we get
\begin{equation}
\label{eq3}
\Pi(k=0)=\frac{g^2 T^2}{24 \pi^3}.
\end{equation}
At the pole of the corrected propagator, we have
\begin{equation}
\label{eq3-1}
k_0^2=\Pi(k=0) \equiv M_1^2.
\end{equation} 
The MSCR dictates that the dressed mass is obtained self-consistently at the pole of the full propagator, in a way that the (temperature dependent) divergences can be absorbed~\cite{Caldas1,Caldas2,Caldas3,Caldas4}. 
\begin{equation}
\label{eq4}
M_2^2= (1+A_2)M_1^2+\Pi(M_1^2,k_0=M_2,k=0),
\end{equation}
where $A_2$ is the coefficient of the appropriate temperature dependent counterterm and $\Pi(M_1^2,k_0,k)$ is expressed as
\begin{equation}
\label{eq5}
\Pi(M_1^2,k_0,k)=\frac{1}{2}\frac{g^2 T}{(2 \pi)^5} \sum_{n} \int_{\lambda}^{T} d^5 p~\frac{1}{w_n^2+p^2 + M_1^2}~\frac{1}{(w_n+w_l)^2+({\bf p}+{\bf k})^2 + M_1^2},
\end{equation}
where $M_1$ is given by Eq. (\ref{eq3-1}). As we showed in Ref. \cite{Caldas1,Caldas2,Caldas3,Caldas4}, the resummation in the MSCR is achieved by the recalculation of the self-energy. This first iteration (recalculation) of the self-energy correspond to a sum of an infinite subset of daisy-type diagrams. Successive recalculations of the self-energy would turn Eq.~(\ref{eq4}) in 
\begin{equation}
\label{eq4-2}
M^2= \Pi(M,k_0=M,k=0).
\end{equation}
This is nothing, but the Schwinger-Dyson equation for the two-point function evaluated at zero momentum at the pole of the dressed propagator.

To obtain the pressure, let us use the relation \cite{Kapusta}:
\begin{equation}
\label{eq6}
\left(\frac{\delta P}{\delta {\cal D}_0} \right)_{{\rm 1PI}}=-\frac{T}{2}\Pi,
\end{equation}
where ${\cal D}_0$ is the free-particle propagator and ${\rm 1PI}$ means that only one-particle-irreducible diagrams enter $\Pi$. In this case, $\Pi(k_0,k)$, as given by Eq. (\ref{eq5}), is the one-loop self-energy plus an infinite subset of diagrams summed nonperturbatively through the MSCR. Then, in this approximation, the pressure is
\begin{equation}
\label{eq7}
P = P_0 -\frac{T}{2} \sum_{l=0} \int_{\lambda}^{T} \frac{d^5 k}{{(2\pi)}^5}~\frac{1}{k^2}~\Pi(M_1^2,k_0,k).
\end{equation}
Let us now evaluate $\Pi(M_1^2,k_0,k)$. As will be clear below, from here now $\lambda$ can be set equal to zero with no harm to the pressure, due the presence of $M_1$. Using again the maximization procedure given by Eq. (\ref{eq2}), we write the self-energy with $k_0=0$ (since $l=0$) as
\begin{eqnarray}
\label{eq7-1}
\Pi(M_1^2,k_0=0,k)= \frac{1}{2} g^2 T \int_{0}^{T} \frac{d^5 p}{{(2\pi)}^5}~ \frac{1}{p^2 + M_1^2}~\frac{1}{({\bf p}+{\bf k})^2 + M_1^2}
\\
\nonumber
=\frac{g^2 T}{24 {\pi}^3} \left[\frac{1}{k^2+M_1^2} \left(\frac{k^3}{3}-M_1^2k+ M_1^3 \arctan\frac{k}{M_1}\right) \right.\\
\nonumber
\left. + T -\frac{3}{2} M_1  \arctan\frac{T}{M_1} + \frac{T}{2} \frac{M_1^2}{M_1^2+T^2} 
-\left(  k -\frac{3}{2} M_1  \arctan\frac{k}{M_1} + \frac{k}{2} \frac{M_1^2}{M_1^2+k^2}\right) \right].
\end{eqnarray}
Plugging the equation above in Eq.~(\ref{eq7}) yields

\begin{eqnarray}
\label{eq8}
P = P_0 -\frac{g^2 T^2}{(24 {\pi}^3)^2} \left[ \frac{T^4}{6}+\frac{T^4}{6} \frac{M_1^2}{M_1^2+T^2}-\frac{7}{6}M_1^2~T^2 + M_1^3~T \arctan\frac{T}{M_1} \right.\\
\nonumber
\left. +  \frac{2}{3}M_1^4 \ln \left(1+ \frac{T^2}{M_1^2}\right)-\frac{M_1^4}{2} \arctan^2\frac{T}{M_1} \right].
\end{eqnarray}
Thus we have that the pressure is finite. Replacing the thermal mass given by Eq. (\ref{eq3-1}) in Eq. (\ref{eq8}), we obtain 

\begin{eqnarray}
\label{PressureMax}
P &=&P_{0}-\frac{g^{2}T^{6}}{6 (24 {\pi}^3)^2} \left[1+\frac{\mu^{2}}{1+\mu ^{2}}-7\mu ^{2}+
6\mu ^{3}\arctan \frac{1}{\mu}
\right.\\
\nonumber
&&\left. + 4 \mu ^{4}\ln \left( 1+\frac{1}{\mu ^{2}}\right) -3 \mu ^4 \arctan^2 \frac{1}{\mu } \right],  
\end{eqnarray}
where 

\begin{equation}
\mu ^{2}=\frac{M_{1}^{2}}{T^{2}}=\frac{g^{2}}{24\pi ^{3}}.
\end{equation}
Equation~(\ref{PressureMax}) shows explicitly that the pressure goes with $T^{6}$, as it must be, by dimensional analysis. The factor between brackets can be evaluated as a function of the coupling $g$. In fact, as the value of the parameter $\mu ^{2}$ is small even for $g=1$, the leading term in Eq.~(\ref{PressureMax}) reads

\begin{equation}
P=P_{0}-\frac{g^{2}T^{6}}{6(24\pi ^{3})^2}.
\end{equation}

\section{Comparison between the maximization procedure and Feynman parametrization}

In order to verify the reliability of the maximization procedure employed, we compare its results with results obtained by using Feynman parametrization and shifting the integration variable in the computation of the self-energy. We must remark that this is also an approximate procedure, since no surface effects are taken into account, and the results are valid only in the high temperature limit. Taking again only the zero frequency terms, the self-energy, Eq.~(\ref{eq1}), after introducing one Feynman parameter, reads

\begin{equation}
\label{autoe}
\Pi (k_{0},k)=\frac{g^{2}T}{24\pi ^{3}}\int_{0}^{1}dx\left\{ T+\frac{1}{2}
\frac{M^{2}T}{T^{2}+M^{2}}-\frac{3}{2}M\arctan \left(\frac{T}{M}\right)\right\}, 
\end{equation}
where $M^2=k^{2}x(1-x)-k_{0}^{2}x$, and the integration in $p$ was already done.

Computing the scalar field thermal mass as before, we obtain at the pole of the corrected propagator

\begin{eqnarray}
k_{0}^{2}&=&\Pi (k_{0},k=0)\\
\nonumber
&=&\frac{g^{2}T^{2}}{24\pi ^{3}} \left\{\frac{3}{2}+\frac{T^{2}}{2k_{0}^{2}}
\ln \left(1-\frac{k_{0}^{2}}{T^{2}}\right)-\frac{3}{2}\int_{0}^{1}dx~\frac{k_0}{T}\sqrt{x}\arctan h\left(\frac{T}{k_{0}\sqrt{x}}\right) \right\}.
\end{eqnarray}
For $k_{0}<<T$, we obtain, at leading order, the same result as in the maximization procedure

\begin{equation}
k_{0}^{2}=\frac{g^{2}T^{2}}{24\pi ^{3}} = M_{1}^{2}.
\end{equation}

To evaluate the pressure we use, in Eq.~(\ref{eq6}), $\Pi (k_{0}=0,k)$ from Eq.~(\ref{autoe}). In this limit, by applying the Feynman parameterization and shifting the momentum variable, we obtain

\begin{eqnarray}
\Pi (k) &=&\frac{g^{2}T^2}{24\pi ^{3}} \left[ 1+\frac{1}{2} \left(1+\frac{4T^{2}}{F(k,T)}\arctan h\left( \frac{k^{2}}{F(k,T)}\right)   \right)
\label{PipFeyn} \right.\\
\nonumber
&&\left.
-\frac{3}{2T}\int_{0}^{1}dx\sqrt{k^{2}x(1-x)+M_{1}^{2}}\arctan \left( \frac{T}{\sqrt{k^{2}x(1-x)+M_{1}^{2}}}\right) \right],  
\end{eqnarray}
where $F(k,T) \equiv \sqrt{k^{4}-4k^{2}(T^{2}+M_{1}^{2})}$.

Figure I compares Eq. (\ref{PipFeyn}) with the result obtained within the maximization procedure, Eq.~(\ref{eq7-1}), for $g=0.1$. We can observe that the behavior of both results are approximately the same, in the region bellow the temperature cut-off.

\begin{figure}[t]
\includegraphics[height=3.0in]{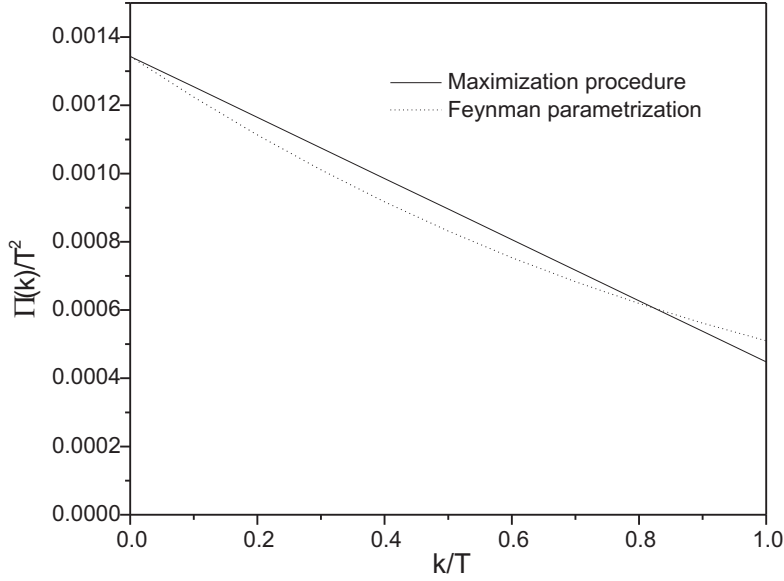}
\caption{\label{1}\textit{  A comparison between the maximization procedure and Feynman parametrization approaches to the self-energy as a function of $k/T$.  } }
\end{figure}

Using Eqs.~(\ref{eq7}) and (\ref{PipFeyn}), we obtain, for the pressure in the Feynman parametrization approach:

\begin{eqnarray}
P &=&P_{0}-\frac{g^{2}T^{6}}{(24\pi ^{3})^{2}}\left[\frac{1}{2}+2\int_{0}^{1}dy\,
\frac{y}{\sqrt{y^{2}-4(1+\mu ^{2})}}\arctan h\left( \frac{y}{\sqrt{
y^{2}-4(1+\mu ^{2})}}\right)   \label{PressureFeyn}  \right.\\
\nonumber
&&\left.
-\frac{3}{2}\int_{0}^{1}dx\int_{0}^{1}dy\,y^{2}\sqrt{y^{2}x(1-x)+\mu ^{2}}
\arctan \left( \frac{1}{\sqrt{y^{2}x(1-x)+\mu ^{2}}}\right) \right].  
\end{eqnarray}

In order to compare Eq.~(\ref{PressureFeyn}) with Eq.~(\ref{PressureMax}), we compute numerically the value of $-(24\pi^{3})^{2}\frac{P-P_{0}}{T^{6}}$, a result that is independent of $T$, in the limits $g= 0.1$ and $g= 1$. Results are shown on Table I.

\[\begin{tabular}{|c|c|c|}
\hline
& $g=0.1$ & $g=1$ \\ \hline
Maximization procedure & 0.00167 & 0.165 \\ \hline
Feynman parameterization & 0.00130 & 0.129 \\ \hline
\end{tabular}\]
TABLE I: Numerical evaluation of $-(24\pi^{3})^{2}\frac{P-P_{0}}{T^{6}}$, from Eqs.~(\ref{PressureMax}) (maximization procedure) and (\ref{PressureFeyn}) (Feynman parametrization), as a function of the coupling $g$, for $g= 0.1$ and $g= 1$.

\section{Conclusions}
\label{conc} 

In this paper we have considered the effects of the resummation in the study of the infrared behavior of the pressure in the $g \phi^3$ scalar theory in 6 dimensions. For that purpose we have used the MSCR, which resums higher-order terms in a nonperturbative fashion. The computation of the scalar self-energy has been done both using a maximization procedure proposed by Almeida and Frenkel as well as the Feynman parametrization. We found that the two approximations agree in the limits analyzed, giving confidence in the procedure we used. We have shown that only one recalculation of the self-energy, which is equivalent to a sum of an infinite subset of diagrams, is enough to ensure the finiteness of the pressure. Although the results for the self-energy and the pressure given by the two approximations used differ a little bit from each other, they ensue a finite expression for these quantities in the infrared limit. 

We note that other approaches like the 2PI\footnote{A study on the renormalization of 2PI effective actions in scalar field theories can be found in Ref.~\cite{van}.}~\cite{Cornwall} and the 2PPI~\cite{Verschelde} (one of its variations) effective actions are also used to perform systematic selective resummations and furnish nonperturbative results such as the MSCR. For some models recently studied in the literature, and under certain approximations, we could point out some similarities between the MSCR and, for instance, the 2PI results. Consider, for example, the 2PI effective potential at finite temperature for the $\lambda \phi^4$ model, $V(\phi,G(k))$, where $G(k)$ is the dressed propagator. After minimizing $V(\phi,G(k))$ with respect to $G(k)$ one obtains a gap equation for $G(k)$. If a Hartree form is taken for the dressed propagator~\cite{Amelino,Nicholas}, $G(k)=1/k^2+M^2$, this yields an equation for the effective mass, which is Eq.~(\ref{eq4-2}) with $\Pi(M,k_0=M,k=0) \to \Pi(M)$ in this case, where $\Pi(M)$ is the bubble diagram~\cite{Caldas3}. However, the 2PI effective resums only the two-point correlation function, leaving the vertex as the bare one leading to gauge dependent results for physical quantities. The gauge dependence of the effective action has been studied by Arrizabalaga and Smit~\cite{Arrizabalaga}, whereas gauge independent formalisms have been discussed by Carrington et. al~\cite{Carrington2} and J. Berges~\cite{Berges}. A modified form of the 2PI effective action has been suggested by Mottola~\cite{Mottola} in order to maintain gauge invariance in QED. Since with the MSCR the vertex can be resummed in the same spirit of the two-point functions, it is expected to preserve gauge invariance as well~\cite{Caldas5}. 

Then we conclude that MSCR could serve as an alternative nonperturbative method to calculate the pressure and other physical quantities besides a Dyson resummation.

\section*{Acknowledgements}
H. C. would like to thank Prof. Josif Frenkel for the correspondence on the subject and for enlightening discussions. H. C. acknowledges full support of the Brazilian agency CNPq.


\end{document}